# Lambda-type sharp rise in the width of Raman and infra-red line shape near the Widom line in super-critical water above its gas-liquid critical temperature

**Tuhin Samanta and Biman Bagchi***

Solid State and Structural Chemistry Unit, Indian Institute of Science, Bangalore – 560012, India

E-mail: bbagchi@sscu.iisc.ernet.in

## *Abstract*

**A lambda-type divergent rise of Raman linewidth of liquid nitrogen near its critical temperature has been a subject of many discussions in the past[1-5]. Here we explore the possibility of such an anomaly in infra-red and Raman spectroscopy of super-critical water (SCW) by varying the density across the Widom line just above its critical temperature. Vibrational phase relaxation is expected to be a sensitive probe of fluid dynamics. We carry out computer simulations of two different model potentials (SPC/E and TIP4P/2005) to obtain the necessary time correlation functions. An additional feature of this work is a quantum chemical calculation of the anharmonicity parameter that largely controls frequency fluctuations. We find a sharp rise in the vibrational relaxation rate (or the line widths) for both the models as we travel across the Widom line. The rise is noticeably less sharp in water than in nitrogen. We attribute this difference to the faster relaxation rate in water. We demonstrate that the anomalous rise is due to sharp increase in the mean square frequency fluctuation and not due to any dynamical critical slowing down because the time constant of the normalized frequency-frequency time correlation functions show lack of any noticeable change as we move across the Widom line. We present an explanation of the observed results using the mode coupling theory of liquid dynamics.**

## I. Introduction

Vibrational phase relaxation, reflected in both Raman and infra-red line shapes, is a sensitive probe of the dynamical state of the system. While studies often focus on unraveling details such as homogeneous versus inhomogeneous contributions to the line width, or the role of anharmonicity of the normal mode whose vibrations are being studied, or to various mechanisms of dephasing, the probe can reveal information about large scale changes occurring in the system. One such example is provided by the work of Musso et al.[1, 2] who has shown that as the critical temperature is approached either along an isochore or an isotherm, the width of the Raman line from liquid nitrogen undergoes a sharp increase. When plotted against temperature and density, the width exhibits a Lambda-type rise which is well known in critical phenomena.

This anomalous behavior of the Raman line width of nitrogen attracted a good amount of theoretical attention[4-6]. Although initially assumed that this sharp rise in relaxation rate is coupled to critical slowing down, it was later demonstrated that the rise is more of a consequence of large scale density fluctuations that appear near the critical temperature. That is, this is a consequence of static heterogeneity.

In this work we are concerned with a similar behavior in the vibrational relaxation of water near its critical temperature. One important difference is that water even in its super-critical state is known to exhibit ultrafast dynamics. It can be of great interest to explore how this ultrafast dynamics of water controls the vibrational relaxation. We note in passing that there have been earlier studies of vibrational relaxation in super-critical water[7, 8], but the study here systematically explores the appearance and disappearance of critical behavior as we cross the Widom line. Also, earlier studies often focused on energy relaxation which is coupled to high frequency modes, as demonstrated by Rey and Hynes[9]. Vibrational energy relaxation is expected to be less sensitive to collective density and energy fluctuations than phase relaxation.

Water just by itself is an interesting object for research because of its great importance in life and in different chemical and biological processes. Apparently simple, it demonstrates an impressive range of unusual properties in the condensed phase. There is probably no other substance that has been studied more thoroughly than water and yet no satisfactory theory exist that can explain its anomalous properties. While there have been many studies in the literatures on various anomalies[1, 2, 4-6, 10-12], relatively fewer studies have focused on super-critical water[13-15]. One expects and finds pronounced anomalies in various response functions (specific heat, isothermal compressibility, diffusion coefficient) as we cross the Widom line in the super-critical state, remaining close to the critical temperature.

Many studies of $H_2O$ have recently focused on the vibrational spectroscopy of the –O-H (and –O-D in HOD) stretching frequency/band. This –O-H stretching frequency of $H_2O$ is known to undergo rapid fluctuations due to fast motions of water molecules and also formation and breakage of hydrogen bonds. For years it is known that the molecules with weak or less number of hydrogen (H) bonds absorb in the blue side of the –O-H stretch, whereas the molecules with strong H-bonds absorb in the red side of the vibrational spectrum. The detailed understanding about this general trend can be perceived by different experimental techniques (such as X-ray, Raman, absorption and emission spectra).

Understanding the structure and dynamics of liquid water has been an important and challenging goal. In this context the Raman and infrared (IR) spectroscopy techniques of the –O-H stretch region have played an important role, since –O-H symmetric stretch frequencies are sensitive to the immediate environments or molecular surroundings and are both IR and Raman active, their spectral characteristics being quite different though. The IR spectrum[16, 17] (at room temperature) peak at ~ 3400 cm$^{-1}$, has a weak shoulder at about 3250 cm$^{-1}$ and a FWHM (full width at half maximum) of about 375 cm$^{-1}$. Raman spectra[18-20] are bimodal in nature, with peaks at about 3400 cm$^{-1}$ and 3250 cm$^{-1}$, and a FWHM of about 425 cm$^{-1}$. Liquid state spectra are significantly red shifted from gas phase spectra and moreover the breadths of liquid state spectra are substantially larger and broader than that in gas phase[16]. Several experimental and theoretical studies have been performed on the IR and Raman lineshape of –O-H/ -O-D stretch[21-27] at room temperature but there have been limited number of studies of lineshape of –O-H stretch in supercritical water. There are relatively fewer studies of water in the super-critical state, except several on vibrational energy relaxation which probes high frequency motions of the liquid.

Previous[4, 28] studies on vibrational phase relaxation of liquid $N_2$ along the critical isochore and critical isotherm reported dramatic increase in dephasing rate near the gas-liquid critical point in the form of a divergence of Raman line width[4, 5]. As already mentioned, this has been explained by invoking dynamical slowing down and the role of static heterogeneity. Additionally, Raman line shape changes from Lorentzian to Gaussian, as we move from temperatures far from critical temperature $T_c$ towards $T_c$, and vice versa.

Widom line is the virtual line that starts at the critical temperature $T_c$ and moves upward. An important aspect of a system (gas or liquid) near the Widom line is its enhanced local density fluctuations. Unlike the critical fluctuations, these can be of shorter ranges but amplitudes can still be large, giving rise to a large amplification in isothermal compressibility ($\kappa_T$) and specific heat ($C_V$). In addition to the thermodynamic anomalies, across the Widom line one is also interested to study the variations in the values of transport properties and relaxation times in order to better understand the origin of the anomalous critical point effects.

Not much is known about the microscopic dynamical properties of these fluctuations in supercritical water (SCW). As many chemical processes occur in supercritical water, it is necessary to obtain a measure of the dynamics of this system. Because of high temperature, the dynamics is expected to be ultrafast. In liquid water at ambient conditions orientational correlation time is of the order of one ps. In SCW, the time scale is expected to be shorter.

In the present work we study two different water models, namely SPC/E and TIP4P/2005. Both temperature dependent and density dependent vibrational dephasing rate show a $\lambda$-shaped rise in values (see **Figure S6 (b)** in Supplementary material and **Figure 4(a)**) across the Widom line. For both models, frequency-frequency time correlation functions (FFCF) show an initial fast drop having a time constant of the order of 30 fs and a slower diffusive component with a time constant of the order of 230 fs, with an average correlation time of the order of 135 fs (see **Table S2** and **Table S3** in Supplementary material). At supercritical state, the average vibrational dephasing time is less than 100 fs (see **Table S2** and **Table S3** in Supplementary material) for both the water models. The main results from the two different water models are in semi-quantitative agreement.

In this report we discuss the theory necessary to study VPR and the vibrational lineshape of –O-H stretch by both Raman and IR spectroscopic methods/techniques. We study the line broadening of the vibrational lineshape and $\lambda$-shaped divergence in Raman and IR linewidths of SCW near the gas-liquid critical point. The characteristic nature of absorption maxima of Raman/IR lineshape as a function of density of SCW is reported here. We use both molecular dynamics simulation and mode coupling theory to analyze and explain the results obtained through simulation. In this study, we also calculate rotational relaxation and dielectric relaxation times across the Widom line of supercritical water.

## II. Anomalies in thermodynamic response functions

Linear response functions (like $C_V$ and $\kappa_T$), expressed as $C_V = \left(\frac{\partial U}{\partial T}\right)_V$ and $\kappa_T = -\frac{1}{V}\left(\frac{\partial V}{\partial P}\right)_T$, are known to diverge as the critical point is approached. On the other hand, a sharp rise is seen in these functions as the Widom line is crossed.

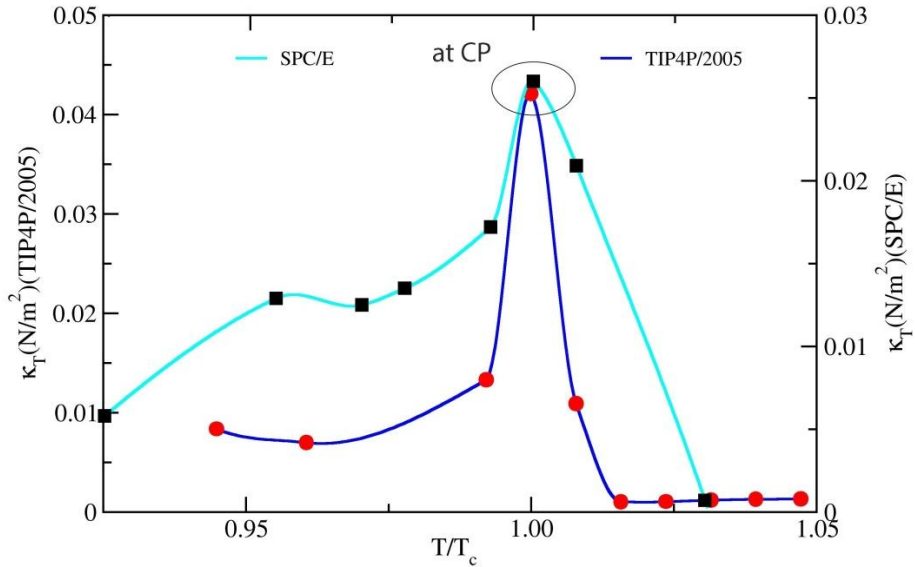

**Fig. 1: The temperature dependence of isothermal compressibility. Note that the thermodynamic response function, isothermal compressibility, $\kappa_T \left(= \frac{<\Delta V^2>}{V K_B T}\right)$, exhibits a sharp maximum at temperatures 660 K (for SPC/E water model) and 635 K (for TIP4P/2005 water model). We scale the temperature of the system with the critical temperature of the corresponding water model. From our calculation, critical temperatures are ( $T_c$ ) = 660 K (for SPC/E) and 635 K (for TIP4P/2005).**

**Figure 1** depicts the marked enhancement in isothermal compressibility ($\kappa_T$) when temperature of water is varied at a constant volume (above the critical pressure), across the critical temperature. A similar anomaly is exhibited by the specific heat, $C_V$, which is documented in the Supplementary material (in **Figure S1**). The calculated $\kappa_T$ and $C_V$ show a sharp maximum precisely at the critical temperature which is 660 K for the simple point charge (SPC/E) and 635 K for TIP4P/2005 water models. We have performed simulations at temperatures 670 K (for SPC/E) and 650 K (for TIP4P/2005) (that is just above the critical temperatures of the consecutive water models) by varying the density. For SPC/E and TIP4P/2005 water models critical densities are respectively 0.33 g/cm$^3$ and 0.31 g/cm$^3$. Near critical point density inhomogeneity is an important phenomenon in SCW, supported by the experimental results, discussed systematically in recent reviews[29, 30].

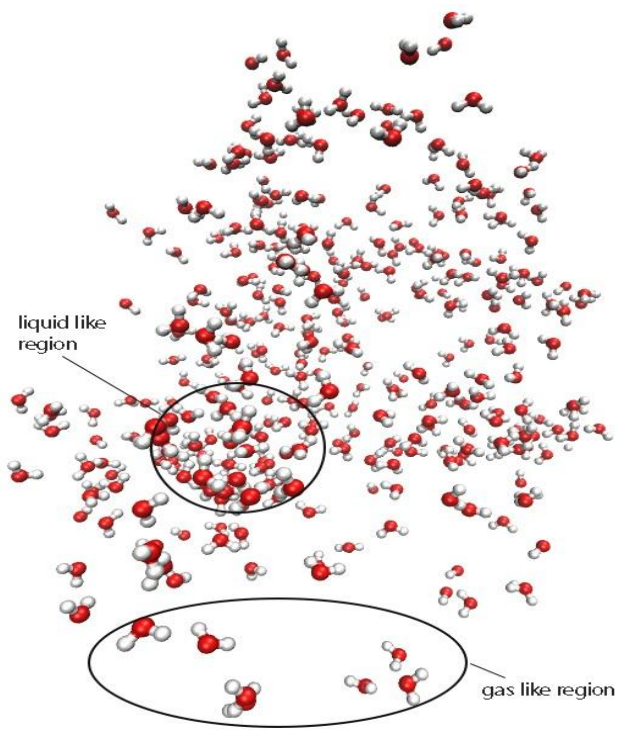

**Fig. 2: Characteristic snapshot of supercritical water (SCW). Near critical point, high density region (liquid –like) and low density region (gas-like) are clearly observed (marked by black circles) in SCW.**

At and just above the critical temperature, there is a coexistence of both gas like and liquid like water domains. In the super-critical water, the system is homogeneous at long length scales, but exhibits density fluctuations of the type shown in **Figure 2**. Such coexistence decreases as we move away from critical temperature but exists just above the critical temperature. In the super-critical temperature region, the boundary between the two regions i.e. liquid-like SCW and vapor-like SCW has been defined by Stanley and his co-workers as the Widom line [31]. In our study we traverse the Widom line by changing the density at a constant temperature (670 K (for SPC/E) and (650 K (for TIP4P/2005)) greater than critical temperature. The simulation details are described in the Appendix.

### III. Theoretical discussions of Raman lineshape

Many experimental studies (IR, Raman line shapes, laser spectroscopy) have been performed to study the vibrational phase relaxation in condensed phase and its dependence on temperature and density[3, 4, 6, 32]. The easiest access to vibrational phase relaxation is the isotropic Raman linewidth measurement. Our understanding of vibrational dephasing[33-36] is based on the Kubo-Oxtoby's[37-40] stochastic theory of lineshape. Taking weak coupling approximation and neglecting the polarizability variation, the isotropic Raman line shape ($I_{iso}$) is determined through the Fourier transform of the normal co-ordinate (Q) time correlation function,

$$I_{iso} = \int_{-\infty}^{\infty} dt\, e^{i\omega t} \langle Q(t)Q(0) \rangle \qquad (1)$$

The experimental observables are either the lineshape function as in the classical experiments or the normal co-ordinate time correlation function. The correlation function of $Q(t)$ is defined by,

$$\langle Q(t)Q(0)\rangle = \text{Re}\exp(i\omega_0 t)\left\langle \exp\left[i\int_0^t dt' \delta\omega(t')\right]\right\rangle \tag{2}$$

In the right hand side of **Eq. 2**, the second term is the cumulant function. We can express the **Eq.2** as,

$$\langle Q(t)Q(0)\rangle = \text{Re}\left[\exp(i\omega_0 t)\right] \times \exp(i\int_0^t dt'\langle\delta\omega(t')\rangle + \frac{i^2}{2}\int_0^t\int_0^{t_1} dt'dt''\langle\delta\omega(t')\delta\omega(t'')\rangle + \frac{i^3}{6}\int_0^t\int_0^{t_1}\int_0^{t_2} dt'dt''dt'''\langle\delta\omega(t')\delta\omega(t'')\delta\omega(t''')\rangle + \ldots\ldots) \tag{3}$$

where the brackets are the cumulant averages.

A cumulant expansion of **Eq. 3** followed by truncation after second order (see section **S.2.I** in Supplementary material) gives the following well-known expression of the normal coordinate time correlation function in terms of frequency frequency time correlation function (FFCF)[40].

$$\langle Q(t)Q(0)\rangle = \text{Re}\left[\exp(i\omega_0 t + i\langle\delta\omega\rangle t)\right] \times \exp\left[-\int_0^t dt'(t-t')\langle\delta\omega(t')\delta\omega(0)\rangle\right] \tag{4}$$

where, $\delta\omega_i(t) = \omega_i(t) - <\omega_i>$ is the difference of the vibrational frequency from average vibrational frequency. $<\delta\omega(0)\delta\omega(t)>$ is the frequency fluctuation time correlation function(FFCF) and $\omega_0$ is fundamental vibrational frequency of the '-O-H' stretch, which is 3500 cm$^{-1}$. Assuming weak coupling of the vibration to the solvent bath, extending the integration limit to the infinity and replacing $t-t'$ by t, one obtains,

$$\langle Q(t)Q(0)\rangle \approx \exp(-t/\tau_V) \tag{5}$$

where,

$$\tau_v^{-1} = \int_0^\infty dt \langle \delta\omega(t)\delta\omega(0)\rangle \tag{6}$$

Therefore the dephasing time is related to the correlation time by the following expression,

$$\tau_v = \frac{1}{\langle \delta\omega^2 \rangle \tau_c} . \tag{7}$$

The correlation time can be defined as,

$$\tau_c = \int_0^\infty \langle \delta\omega(0)\delta\omega(t)\rangle dt / \langle \delta\omega(0)\delta\omega(0)\rangle \tag{8}$$

where $\langle \delta\omega^2 \rangle^{1/2}$ measures the amplitude of modulation.

Oxtoby[38, 41] derived the final expression for the frequency frequency time correlation function (FFCF) between two quantum levels $n$ and 0 as follows,

$$\langle \delta\omega_{n0}(t)\delta\omega_{n0}(0)\rangle = \frac{n^2}{2}\sum_i \left[\frac{3(-K_{111})l_{ik}}{\omega_0^3 m_i^{1/2}} + \frac{l_{ik}^2}{2\omega_0 L m_i}\right]^2 \times \langle F_i(t)F_i(0)\rangle \tag{9}$$

where $K_{111}$ is the anharmonic force constant and $\langle F_i(t)F_i(0)\rangle$ represents the force-force time correlation function on the $i$ th atom moving along the direction of vibration. In Supplementary materials (see section **S.2.II**), we provide the theoretical descriptions of **Eq. 9** in details.

In an elegant series of papers, Skinner and co-worker[42, 43] directly calculated $<\delta\omega(0)\delta\omega(t)>$ from each trajectory. The main objective is to consider the slightly different forces experienced by the bonds in the vibrational excited states. Effects of interactions between the chemical bond and surrounding solvent molecules are manifested in the present formalism through two terms - the anharmonicity term $K_{111}$ and the force term, $F(t)$. For the calculations presented here, the anharmonicity, $K_{111}$, which is the third order expansion term of the potential energy surface, is related to the anharmonic frequency shift ($\Delta$), of the –O-H vibration by the following expression,

$$K_{111} = \frac{\omega_0^2 \sqrt{\Delta}}{2\hbar^{3/2} \sqrt{\mu}} \propto \sqrt{\Delta} \qquad (10)$$

It is well-known that this simplification comes with the possible penalty that $\Delta$ varies for various configurations, along with the force term. For our calculations, the average anharmonic frequency shift value is $\approx 260$ cm$^{-1}$ (see **Figure S4** in Supplementary material). Details of the calculations (ES/MD) can be found in the Supplementary materials (see section **S.2.III** in Supplementary material). The mean square force term, however, has a wider variation, as expected for a system near the critical temperature, leading to a large value of the mean square frequency fluctuation. Variation of mean square frequency fluctuation ($<\delta\omega^2>$) is more sensitive to the $F(t)$, than anharmonicity term.

## IV. Theory of Infrared lineshape

The far-infrared and mid-infrared (2.5-10 μm) regions are the most informative spectral regions for the water molecules as inter and intra-molecular interactions of the system are reflected in the above mentioned frequency ranges. In our case 3 μm region is the most important due to the vibrational band of –O-H stretch frequency, as much dynamical and structural information is hidden in the vibrational part of the water spectrum. The absorption lineshape of -O-H stretch of water has been extensively studied by linear spectroscopy[16, 21, 22, 25, 26, 44]. In the study of infrared spectroscopy of water non-Condon effects can be rather important. The linear absorption line shape is obtained from Fourier transform of linear response function, which can be written, within the semi classical approximation[43, 45] as,

$$I(\omega) \sim \text{Re}\left[\int_{-\infty}^{\infty} dt\, e^{i\omega t} \left\langle \mu_{01}(t)\mu_{01}(0) \times \exp\left[-i\int_{0}^{t} d\tau\, \omega_{01}(\tau)\right]\right\rangle\right] \quad (11)$$

where, $\omega_{01}(t)$ is the time dependent transition frequency of $1\leftarrow 0$ and $\mu_{01}(t)$ is the 1-0 matrix element of the dipole moment which is also time dependent. We define, $\delta\omega_{01}(t) = \omega_{01}(t) - \langle\omega_{01}\rangle$ and neglect the orientational dynamics implicit in **Eq.11** to obtain

$$I(\omega) \sim \text{Re}\left[\int_{-\infty}^{\infty} dt\, e^{i(\omega-\langle\omega_{01}\rangle)t} \left\langle \mu_{01}(t)\mu_{01}(0) \times \exp\left[-i\int_{0}^{t} d\tau\, \delta\omega_{01}(\tau)\right]\right\rangle\right]. \quad (12)$$

We explicitly consider the variation of –O-H transition dipoles with frequency (non-Condon effects) by allowing the variation of $\mu_{01}$ with time. The –O-H transition dipole variation has been demonstrated to play a strong role in the vibrational spectroscopy of water[42]. The transition dipoles and transition frequencies of –O-H stretch of water are calculated from each trajectory.

Slightly different forces are experienced by the bonds due to immediate solvent environments. The fluctuating transition dipoles and frequencies are calculated for the –O-H stretch of $H_2O$ by the combination of electronic structure (ES) calculations and molecular dynamics (MD) simulations, as discussed in detail previously[43, 45].

In ES/MD approach the first step is to extract the random clusters (different snaps with a time interval of 1 ps) of the water molecules and its local environment from an MD simulation of the system. Molecules within 4.5 angstrom have their positions recorded; these nuclei will be included as the point charges in the following ES calculations. We perform the ES calculation on the clusters using the GAUSSIAN09 software package. These clusters are considered to form representative sample of the environments the particular water molecule encountered in systems. This ES/MD procedure yields $\mu_{01}$ and $\omega_{01}$ of –O-H bond of a water molecule in each of the 100 clusters. The next step of the ES/MD approach is to calculate the electric field along –O-H bond, at a site of H atom, due to water molecules in the clusters. In the supplementary material we provide method for determining the trajectories of required fluctuating quantities from an MD simulation in details (see section **S3** in Supplementary material).

## V. Simulation results of Raman and infrared lineshapes

Both vibrational phase relaxation (VPR) and vibrational energy relaxation (VER) near the gas-liquid critical point and above the critical temperature in the supercritical state have drawn attention in the past[1, 3]. These include different explanations for the divergence like growth of the Raman[3-5, 29] and infrared line widths. In particular, there remains an unsolved controversy as to the precise origin of the critical anomaly. One school of thought advocates the role of critical slowing down[46] of dynamics, while second school of thought[5, 28] suggests that it is predominantly the rise in static heterogeneity that gives rise to the anomalous growth in the dephasing rate as we approach the critical region. The above mentioned controversy provided additional motivation to examine the origin of anomalous vibrational relaxation. Theoretical and computer simulation studies have attempted to explain this $\lambda$-shaped temperature (or density) dependence by appealing to the enhanced density fluctuations near the critical point. The measuerment of both line shift and linewidth is essential to obtain the information about the structrure and dynamics of liquids.

The majority of vibrational dephasing studies have looked into the nature and width of the isotropic Raman lineshape. The Raman lineshape is nothing but the Fourier transform of the coherence decay that characterizes dephasing. In our calculation, Raman lineshape of the –O-H stretch of SCW is calculated by using **Eq. 1** (as in section III). In **Figure S10** (see Supplementary material) we show the theroretical Raman lineshape for differenet densities across the Widom line. There is a crossover observed from the Lorentzian-like to a Gaussian–like lineshape[4, 5, 28], as the critical point is approached, which is in good agreement with previous results for superitical nitrogen[2-5].

On the other hand, the linear absorption spectrum (infrared) of the –O-H stretch of SCW is calculated by using **Eq. 11** (as in section IV). This absorption lineshape is sensitive to the frequency distributions, which characterize the solvation environment present in the system. **Figure S11** (see Supplementary material) shows the calculated theoretical infrared (IR) lineshape for different densities across the Widom line.

The above discussed lineshapes of water molecules in liquid phase get red shifted and broadened compared to the same in gas phase. This arises due to the several interactions (such as H-bonding) among water molecules in the liquid phase. Results obtained from our calculations are in good agreement with the previous theoretical and experimental results of supercritical fluids[2-5, 32, 47-50].

Now, we are interested to know about the cause of ultrafast dephasing of super-critical water with the help of computer simulations. Substantial broadening and complexity of the shape of the spectral bands are also noticed near the gas-liquid co-existence line that can be understood by several analyses.

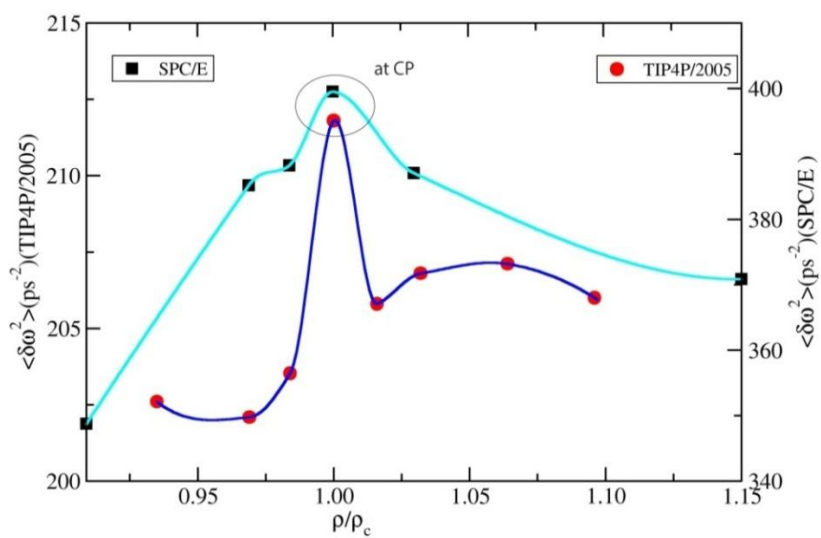

**Fig. 3: Mean square frequency fluctuation ($<\delta\omega^2>$) of SPC/E and TIP4P/2005 water models is plotted against scaled density, at constant temperature (670K for SPC/E water model and 650K for TIP4P/2005 water model). We scale the system density with the critical density of the corresponding water model ($\rho_c$ = 0.33 g/cm³ for SPC/E and $\rho_c$ = 0.31 g/cm³ for TIP4P/2005).**

The calculated frequency fluctuations (see **Figure 3** and **Figure S6 (a)** in Supplementary material), as discussed in section III, show a non-monotonic behavior with density (for both SPC/E and TIP4P/2005 water models), temperature (for SPC/E water model) and display a maximum at the critical point. This signifies increase in the density fluctuations near the critical point.

The exponential decay described in **Eq. 5** gives rise to a Lorentzian linewidth with a half width at half maxima (HWHM) of $\tau_v^{-1}$. The rate of dephasing can be obtained by combining the **Eqs. 4** and **9**, and making a Markovian (exponential) approximation for the decay of frequency-frequency time correlation function (FFCF). In the description of phase relaxation, the assumption of the exponential decay (in **Eq. 5**) allows us to determine the dephasing rate from **Eq. 6**, and provides a relation between dephasing time and frequency-frequency time correlation function (FFCF) in a concise way( see **Eq. 7**).

The simulated Raman and infrared line widths i.e. the full width at half maximum (FWHM) of SCW, obtained from previously discussed lineshapes are plotted as a function of density, shown in **Figure 4 (b).** Near gas-liquid critical point simulated density dependent dephasing rate (see **Figure 4(a)**) or linewidth (see **Figure 4 (b)**) of SCW shows $\lambda$ -type[3-5, 47] transition similar to the experimental results observed by Musso et al.[1, 2, 51].

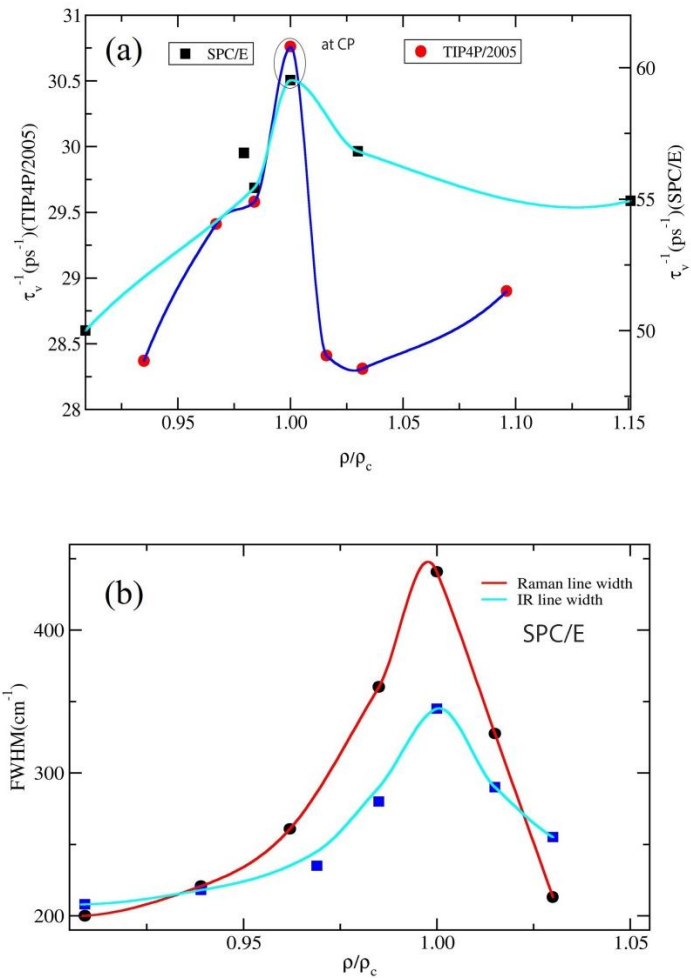

**Fig. 4: (a) Vibrational dephasing rate ($\frac{1}{\tau_v}$) of SPC/E and TIP4P/2005 water models is plotted against scaled density, at constant temperature (670K for SPC/E water model and 650K for TIP4P/2005 water model). We scale the system density with the critical density of the corresponding water model. (b) Theoretical Raman and IR line widths (FWHM) are plotted against density (SPC/E water model). The line width diverges near the critical point. Density of the system is scaled with the critical density, 0.33 g/cm$^3$.**

This $\lambda$-type rise in the line width (or dephasing rate) is due to enhanced density fluctuations near the critical point. The contribution to the density term reflects the combined effects of collisions, attractive and repulsive interactions on the friction on the bond. These terms naturally grow as the critical point (CP) is approached. Thus, as we approach the critical point, frequency fluctuation is observed due to large density inhomogeneity in the system. Local inhomogeneity[2, 4] is playing the key role to affect resultant dynamical behavior near critical point.

The peak maxima of Raman and IR line shape of –O-H stretch are plotted here as a function of density, shown in **Figure 5**. It is clearly a red shift of a few wave numbers. The magnitude of line shift increases with density as clearly seen in **Figure 5**.

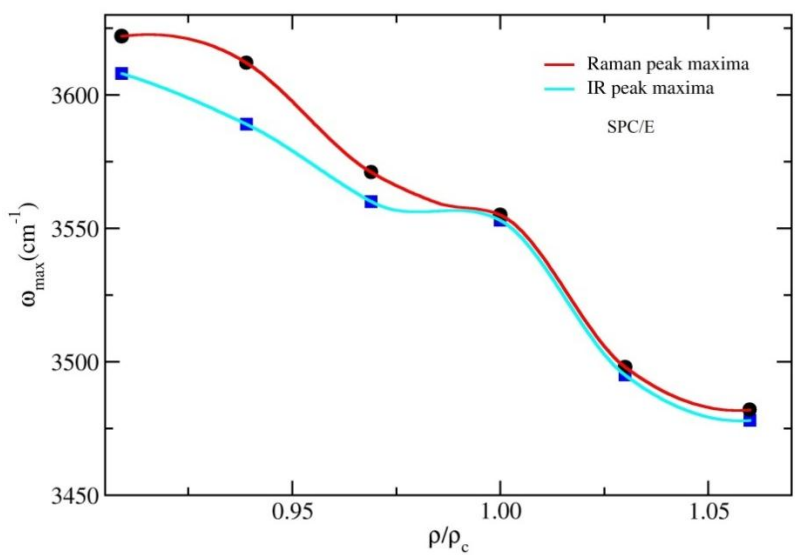

**Fig. 5: Peak maxima positions of the Raman and IR line shape are plotted with scaled densities. Peak positions are gradually shifted towards the lower frequency value with increasing density.**

Gradual change in the positions of absorption maxima on going from the gas like region to the liquid like region is attributed due to the forces in the condensed phase medium. These forces shift the vibrational frequency towards the red side of the frequency axis. Near critical point, peak positions are density independent which has also been experimentally observed for supercritical fluids[32, 49, 50]. The red shift in the IR line shape relative to the Raman lineshape is due to the non-Condon effects.

## VI. Rotational and dielectric relaxation across the Widom line

Several prior studies were devoted to understand the reorientational dynamics of supercritical water[14, 52, 53]. Since infra-red spectrum can be influenced by orientational motion of the bond in question, it is necessary to explore orientational dynamics as well. An additional issue is that critical behavior is reflected more strongly in collective properties than in single particle properties. We thus study dielectric relaxation and report here.

Time correlation function is generally used to analyze the molecular reorientational correlation motion in liquids and is given by[54] as follows,

$$C_{l,\alpha} = <P_l(\vec{u}_\alpha(t)\vec{u}_\alpha(0))> \tag{13}$$

$\vec{u}_\alpha$ is the unit vector along the z direction, $\vec{u}_{OH}$ is along the -O-H direction and $P_l$ refers the $l^{th}$ order Legendre polynomial. Experimentally, rank 1 orientational correlation functions ($C_{1,OH}$) are related to the dielectric measurement and rank 2 orientational correlation functions ($C_{2,OH}$) are related to the NMR/IR relaxation measurement.

Rotational correlation function shows bimodal relaxation similar to previous findings[13, 54] (see **Figures S12** and **S13** in Supplementary material). Thus, the orientational relaxation reported here suggests that there are two processes involved in the molecular rotations, one of them is due to weakly hydrogen bonded water molecules and the other due to strongly hydrogen bonded water molecules. For weakly hydrogen bonded water molecules (blue shifted -O-H bond stretching frequency) orientational relaxation is much faster than the strongly hydrogen bonded water molecules (red shifted -O-H bond stretching frequency), where orientational relaxation time is slow (see **Table S4** and **Table S5** in Supplementary material). For water molecules at high temperature the extensive hydrogen bond network is completely destroyed, thereby eliminating the contribution from librational and intermolecular vibrational modes. The ultrafast component arises due to fast rotational motion of small water molecules[14].

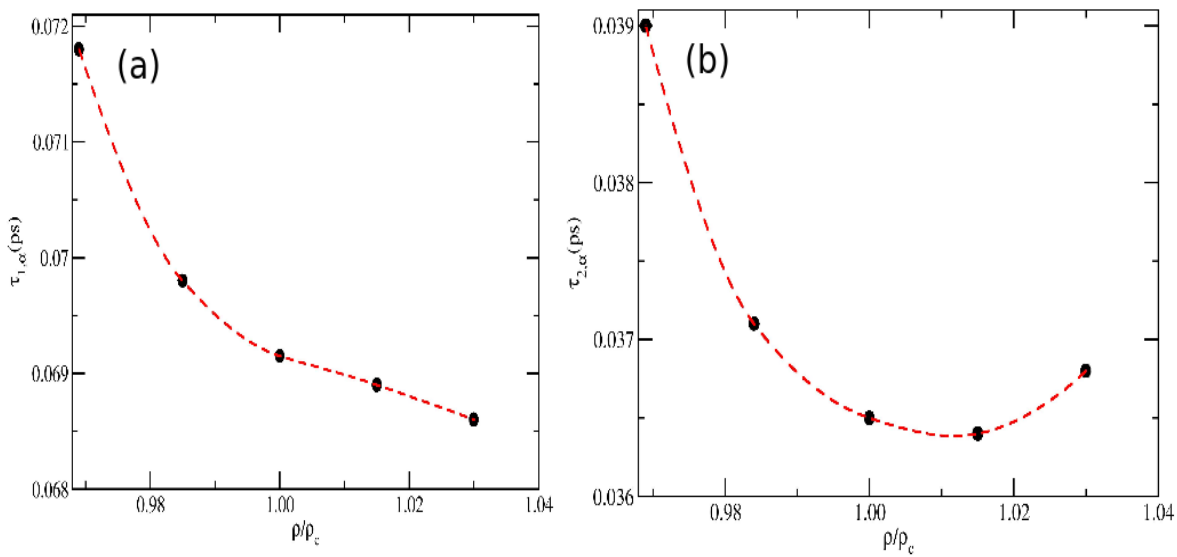

**Fig. 6: Variation of orientational correlation times ($\tau_{l,\alpha} = \int_0^\infty C_{l,\alpha}(t)dt$) with density for (a) 1$^{st}$ rank and (b) 2$^{nd}$ rank correlation time of supercritical water.**

From **Figure 6** we can observe the orientational correlation times (both 1$^{st}$ rank and 2$^{nd}$ rank correlation time) for different densities across the Widom line. Near critical point no significant slowing down of the orientational relaxation is observed.

Among several experimental techniques, the study of dielectric relaxation (DR) has proved to be rather fruitful for critical understanding about water dynamics. The dielectric relaxation measures the collective polarization of all the polar molecules present in the system[55, 56]. The dielectric relaxation time usually allows a measure of the time taken by a system to reach the equilibrium (final) polarization in the sudden on (or off) of an external field. Frequency dependent dielectric constant, ε (ω) is related to the total dipole time correlation function $\langle M(0)M(t)\rangle$ (see **Figure S14** in Supplementary material). *M(t)* is the sum of the all individual molecule dipole moments, $\vec{\mu}_i(t)$: $M(t) = \sum_i \vec{\mu}_i(t)$.

The dielectric spectra of bulk water in the Debye relaxation region is traditionally characterized by two relaxation times with time constants 8.2 and 1.02 ps respectively[55-57]. However for supercritical water, relaxation is expected to be fast compared to bulk water (see **Figure 7**). To quantify the contribution of water dipoles to the system's dielectric constant, we plot the normalized total dipole-dipole time correlation, ($\langle M(t)M(0)\rangle$) (see **Figure S14** in Supplementary material).

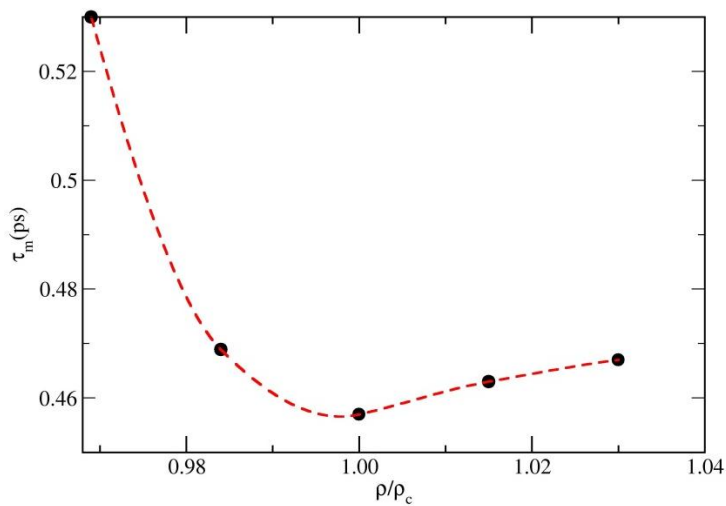

**Fig. 7: Dipolar correlation times ($\tau_M = \int_0^\infty \Phi_M(t)dt$) are plotted against scaled densities of supercritical water. Anomalous drop in the dipolar relaxation time is noticed around critical density. Dynamics is fast near the critical point.**

When fitted to the exponential forms, $\langle M(t)M(0)\rangle$ gives a bi-exponential fit. The amplitudes and time constants are given in the supplementary (see **Table S6** in Supplementary material).

The first component, having time constant less than 100 fs with small amplitude, is the signature of free water molecules that have faster dynamics. The slower component of the dipolar relaxation is of the order of 400 fs with larger amplitude, essentially characterizing the existence of hydrogen bonded water molecules. The average relaxation time shows a minimum which implies an acceleration of dielectric relaxation.

At room temperature the dielectric constant[55] value of SPC/E water model is ~ 68, which is also reproduced here. At high temperatures, the static dielectric constant value is low and lies in between 6-8 (see **Table S7** in Supplementary material). This diminution of the dielectric constant value of SCW is also reported in literature[58, 59].

## VII. A mode coupling theoretic explanation of anomaly

As the density of SCW is varied across the Widom line, neither self-diffusion coefficient nor the rotational correlation time shows any anomaly across the Widom line. The total dipole moment time correlation, $\langle M(t)M(0) \rangle$, demonstrates a weak anomaly. In contrast, the vibrational phase relaxation exhibits pronounced anomaly similar to the ones exhibited by the specific heat and isothermal compressibility as we cross the Widom line. These contrasting behaviors can be easily understood from the mode coupling theory (MCT) analysis of vibrational relaxation described in Ref.[60].

We first note that frequency-frequency time correlation function, $<\delta\omega(0)\delta\omega(t)>$, is related to the energy-energy time correlation function by the following simple relation,

$$<\delta\omega(0)\delta\omega(t)> = \hbar^2 <\delta E_N(0)\delta E_N(t)> \qquad (14)$$

Where $\delta E_N(t)$ is the fluctuation in the energy of the normal mode whose phase relaxation is being studied. A combination of time dependent density functional theory (DFT) and mode coupling theory (MCT) analyses leads to the following expression for the dominant contribution to the frequency-frequency time correlation function.

$$<\delta\omega(0)\delta\omega(t)> = \frac{(k_B T)^2 \rho}{6\pi^2 \hbar^2} \int_0^\infty dq\, q^2\, C_{NO}^2(q) F_{NS}(q,t) F_{OO}(q,t) \qquad (15)$$

where, $C_{NO}(q)$ is the wave number ($q$) dependent direct correlation function between the normal mode and the solvent molecules. $F_{NS}(q,t)$ is the self-dynamic structure factor of the normal mode and $F_{OO}(q,t)$ is the dynamic structure factor of the supercritical water. **Eq. 15** is approximate but captures the main features as discussed below.

In **Eq. 15**, $F_{OO}(q,t)$ exhibits a slow dynamics in the long wave length (q→0) limit. The same time mean square fluctuation, $<\delta\omega^2(0)>$, derives large contribution as the Widom line is approached because,

$$<\delta\omega(0)\delta\omega(t)> = \frac{(k_B T)^2 \rho}{6\pi^2 \hbar^2} \int_0^\infty dq\, q^2\, C_{NO}^2(q) F_{NS}(q) S(q) \qquad (16)$$

where, the static structure factor $S(q)$ exhibis a divergent like growth at small wavenumber since,

$$S(q=0) = \rho k_B T \kappa_T \qquad (17)$$

Neither $C_{NO}(q)$ nor $F_{NS}(q)$ has any divergence–like behavior. Since rate of phase relaxation remains in the ps range and dynamics of super-critical water remains ultrafast, it is $<\delta\omega^2(0)>$ that controls Raman and IR linewidths. Note that MCT analysis shows that neither the self-diffusion coefficient nor the rotational correlation time is expected to exhibit any noticeable slowing down, in agreement with the simulation results.

## VIII.      Conclusions

Several anomalies of supercritical fluids can be understood by using the idea of Widom line. We have combined molecular-dynamics simulations of molecular motions with quantum chemical calculations of anharmonicity to obtain vibrational line widths (both IR and Raman line shapes) across the Widom line. Below we summarize the main results of this work.

i) We employed two different model potentials for water, SPC/E and TIP4P/2005 to study vibration phase relaxation (VPR) of super-critical water across the Widom line.

ii) Both the line shapes show sharp changes for both the potential models as we cross the Widom line by varying the density keeping temperature fixed slightly above the critical temperature of water.

iii) A divergence like behavior of the line width (equivalent to the vibrational phase relaxation (VPR) rate) is observed as we cross the Widom line near but above the critical temperature. The rise in the rate exhibits a $\lambda-$ shaped density dependence of the Raman and IR linewidths (or VPR rate).

iv) A crossover from the Lorentzian–like to a Gaussian-like line shape has been observed as the critical point is approached. This is in agreement with earlier studies on nitrogen[1-5] . However, the amplitude of critical effects is smaller for water. We attributed this to faster dynamics of water molecules. In the case of fluid nitrogen[4, 58] (N$_2$), VPR rate becomes considerably faster (with a sharp peak of the Raman line width) near the critical point (CP) and in this case static  heterogeneity controls the Raman linewidth. In supercritical water, the increase in relaxation rate on approaching the Widom line is found to be *smaller* than that observed for $N_2$. Compared to $N_2$, the magnitude of the correlation time ($\tau_c$) is substantially less for the –O-H bond of water molecules. This effectively explains the slower rise in the VPR rate of supercritical water than that in $N_2$ molecule.

We conclude by noting that at the critical point, there are many interesting dynamical phenomena that occur with substantially smaller time scales than usually discussed in literature. There have been discussions on the time scales of chemical reactions near the critical region[61] in the past.

# Appendix
## Simulation details

Simulations were carried out using 2048 SPC/E water[62] and TIP4P/2005 water[63] molecules in a cubic box. We used GROMACS, version 4.5.5 for MD simulations[64]. Steepest descent method was used for energy minimization of the initial configuration. We used fourth order Particle Mesh Ewald (PME) summation for long range interaction with grid spacing of 0.16. Equations of motion were integrated with time steps of 1 fs during equilibration as well as production runs. We equilibrate the system initially in NVT ensemble over 10 ns. Finally data acquisitions were performed in NVT ensemble. The temperature was kept constant at 670K (for SPC/E) and 650K (for TIP4P/2005) using Berendsen[65] thermostat. Periodic boundary conditions (PBC) were applied in all three directions. We traversed the Widom line by changing the system volume (density) across the constant temperature line.

## Acknowledgements

This work was supported in parts by grants from DST and Sir JC fellowship (DST). We thank Dr. Sarmistha Sarkar and Mr. Saumyak Mukherjee for helping during preparation of paper. T.S. acknowledges CSIR-UGC for providing Senior Research Fellowship (SRF).